\documentclass[12pt]{article}
\begin{document}

\title{\bf The Relativistic Generalization of the Gravitational Force
for Arbitrary Spacetimes}
\author{Asghar Qadir and M. Sharif\\
Department of Mathematics, Quaid-i-Azam University\\
Islamabad, Pakistan}

\date{}

\maketitle

\begin{abstract}
It has been suggested that re-expressing relativity in terms of
forces could provide fresh insights. The formalism developed for
this purpose only applied to static, or conformally static,
space-times. Here we extend it to arbitrary space-times. It is hoped
that this formalism may lead to a workable definition of mass and
energy in relativity.
\end{abstract}
 Unified field theories and other theories of
gravitation.
\section{Introduction}

Despite the elegance of Einstein's geometrodynamics [1] the concept
of "force" is still used extensively. This is probably due largely
to the inertia involved in the radical change of the concepts
required. Accepting the need to maintain the contact with the
earlier concepts, it was argued [2,3] that it would be worthwhile to
re-introduce the force concept into General Relativity (GR). At the
very least it should provide a fresh way of looking at the
consequences of the theory and new insights into its working.
Hopefully it could lead to new tests of GR which had not been
suggested by the purely geometrical formulation. In fact this
re-expression has provided a further understanding of the dynamics
of a neutral test particle in the field of a charged, rotating,
massive point source [4,5]. It has also been used to suggest
possible relativistic explanations of some hitherto unexplained
astrophysical phenomena [6].

There is a more compelling reason to consider forces in relativity.
As has been argued earlier [7], it may provide a way to avoid the
necessity of going to higher (than four) dimensions in an attempt to
unify the forces of nature. A geometrical unification must
necessarily enlarge space-time so that the special role of the usual
four-metric only appears in a projective theory but it (the
four-metric) occurs on a par with other fields in the full theory.
Alternatively gravity must be treated as just another field which
happens to be of spin two and happens to have an Einstein-Hilbert
Lagrangian. In terms of the relativistic equations of motion,
gravity appears on the left-hand side of the equation (in the
Christoffel symbol) while all the other forces appear on the
right-hand side. Either one enlarges the space to include the other
fields in the metric (and the Christoffel symbols) or one can remove
gravity from the left-hand side and display it explicitly on the
right-hand side to see how it can be related to the other forces. Of
course this is not an attempt at a final solution of the problem,
but rather an attempt to find some "signposts" for an alternative
way to tackle the problem.

There is another hope for this approach. The force will be defined
operationally. It will take a simple form in some particular frame
of reference. A sequence of space-like hypersurfaces defined by
these frames of reference would provide a physical basis for a $3+1$
split of space-time. The hope is that this split may provide a
suitable basis for a canonical quantization attempt.

The gravitational force is extracted from  the left-hand side of the
equation of motion by determining the "force" that gives the same
"bent path" in a "flattened out" background space-time as would be
given by the curvature of space-time. The relativistic analogue of
the gravitational force has been called the pseudo-Newtonian ($\psi
N$) force. It was originally calculated for the Schwarzschild and
Reissner-Nordstrom geometries and an attempt was made to deal with
the Kerr-Newmann metric [2]. The former already gives
"electro-gravitic unification" in that it predicts a repulsive force
of a charged source on neutral matter [4]. Though very different
from the current paradigm of unification, this effect is no less a
physical manifestation of unification, at least in principle, than
the''mixing of photons and weak neutral bosons'' in the electro-weak
theory of Glashow-Salam-Weinberg [8]. The extension to other static
metrics was achieved later [3]. The formalism yielded a single
''gravitational potential'', in a special frame of reference, which
turned out to be the conjectured potential mentioned by Hawking and
Ellis [9].

The restriction to static space-times is easily understood.
Classically, force is related to energy, and energy is a conserved
quantity only if there exists a timelike isometry. This restriction
does not allow any time dependence of the $\psi N$ force. Thus there
remains a problem of definition of energy for arbitrary space-times.
Of course there can be no hope of constructing a realistic field
theory based on static space-times. In particular, there would then
be no canonically conjugate "momentum' field. Attempts to extend to
conformally static space-times, while successful [10], were not very
fruitful as, in effect, the time dependence was eliminated there.

To pursue such a programme, it is necessary to obtain an expression
for the extremal tidal force. This extremal tidal force will lead to
an expression for the extended $\psi N(e\psi N)$ force four-vector.
The zero component of the force four-vector will represent the rate
of change of energy of the particle due to a change in the
gravitational field. Usually we can write the force as the gradient
of a scalar potential. Here we obtain two scalar quantities
corresponding to this $e\psi N$ force. The quantity which
corresponds to the time component of the force will give the
potential energy of the test particle which contributes to its time
variation.

The plan of the paper is as follows. In the next section we briefly
review the relevant aspects of the $\psi N$-formalism for the
purposes of application. In sect. \textbf{3} we discuss the extrema
of the general tidal force. In the next section we calculate the
general $\psi N$ force. In sect. \textbf{5} we obtain the extension
of the $\psi N$ potential to arbitrary space-times. In sect.
\textbf{6} we discuss two cosmological examples and finally, in the
last section, we summarize and discuss our results.

\section{The Pseudo-Newtonian Formalism}

Though the gravitational force is not detectable in a freely falling
frame (FFF) \emph{that is so only at a point}, it is detectable over
a finite spatial extent as the tidal force. For example, it could be
measured by an "accelerometer" consisting [2] of two masses
connected by a spring ending in a needle which moves on one of the
masses marked off as a dial. Stretching or squashing the spring
causes the needle to move one side of the zero position or the
other. The usual tidal force would cause the spring to stretch.
However, a repulsion would result in squashing. Thus this
accelerometer could measure and identify attraction and repulsion.

Mathematically, the tidal acceleration is given by
\begin{equation}
A^{\mu}=R^{\mu}_{\nu\rho\pi}t^{\nu}l^{\rho}t^{\pi},\quad(\mu,~\nu,...=0,...,3),
\end{equation}
where $\textbf{R}$ is the Riemann tensor, $\textbf{t}$ a timelike
vector and $l$ is the spacelike "separation" vector representing the
accelerometer. In its rest frame the tidal force on a test particle
of mass $m$ was taken to be
\begin{equation}
F^{\mu}_{T}=mR^{\mu}_{0\nu 0}l^{\nu}.
\end{equation}
The external values of the tidal force can be obtained by requiring
that $\textbf{l}$ be an eigenvector of $\textbf{R}$ in eq.(2). Since
$\textbf{l}$ has no time component in the chosen frame, neither does
$\textbf{F}^*_T$, the extremal value of the tidal force. Physically
the extremal value can be obtained by turning the accelerometer
about, till it gives the maximal reading, and noting the direction
given by it.

For the Schwarzschild metric the relevant principal direction is the
radial direction. Modulo a local Lorentz factor,f eq. (2) gives the
usual Newtonian tidal force for $\textbf{F}^*_T$. The effect of
introducing a charge is to reduce the tidal force by $mQ^2l/r^3$, in
gravitational units. If we also include rotation the principal
direction is no longer radial in general, but lies in the radial
polar ($\phi=constant$) plane. Its angle of inclination to the
radial direction depends on the radial and polar coordinates of the
test particle. Generally eq.(2) yields a cubic equation for the
eigenvalue (which is essentially $|\textbf{F}^*_{T}|)$. This will
always have at least one real root.

If the metric is of Carter's "circular" form $[11]$, or can be
otherwise broken into blocks so that there are isometries in the
block containing the time component [3], eq.(2) can be reduced to
the form of a directional derivative along $\textbf{l}$, by using
Riemann normal coordinates (RNCs) [1],
\begin{equation}
F^i_T=ml.\nabla \Gamma^i_{00},\quad(i,j,...=1,2,3).
\end{equation}
Thus, up to an "integration constant" $m\Gamma^i_{00}$ must give the
$\psi N$ force, $F^i$. This term is fixed by reuiring that there be
no $\psi N$ force in a Minkowski space. The $\psi N$ force is now
\begin{equation}
F^i=m\Gamma^i_{00}.
\end{equation}
We can then write this force as the gradient of a scalar $\psi N$
potential, V. Requiring that the potential also be zero in a
Minkowski space gives
\begin{equation}
F_i=-V_{,i}=\frac{1}{2}m(1-g_{00})_{,i}.
\end{equation}
That this should be the value of $V$ is the conjecture mentioned
earlier [9].

For the Schwarzschild metric $F_i$ is simply the Newtonian
gravitational force for a point mass and the $\psi N$ potential the
usual Newtonian gravitational potential. The inclusion of a charge
introduces a repulsive component of the gravitational force [2]. The
entire structure of the $\psi N$ force for the Kerr-Newmann metric
may be seen in embryonic form in the $\psi N$ potential.
\begin{equation}
V=-m(2Mr-Q^2)/2(r^2+\alpha^2~cos^2\vartheta)
\end{equation}
where $Q$ is the charge and $\alpha$ the angular momentum per unit
mass of the gravitational source. Clearly there will be a polar
component of the $\psi N$ force since $V$ is $\vartheta$ dependent.

\section{The Extermal Tidal Force}

In developing the $\psi N$-formalism $\textbf{t}$ was identified
with the timelike Killing vector so as to provide an easily
integrable expression for the tidal force. Despite the necessity of
staticity for energy conservation, and hence the usual force
concept, the accelerometer would still show a deflection in a
nonstatic situation. As such there must be a "force" embedded in the
geometry. For example, in a Friedmann-model universe the
accelerometer needle must show this deflection over sufficiently
long periods of time and the model must, therefore, have a "force"
in it. This will be so regardless of the fact that there is no
apparent gravitational source in the model. The deflection, which
the needle of the accelerometer indicates, would be the same for all
orientations of the accelerometer. This deflection would, therefore,
be attributed to an expansion of the universe as a whole. It comes
from the gravitational field and not from any gravitational source.
The requirement is to determine the extremal value of the tidal
force for an arbitrary space-time without any time isometry.

To study the consequences of time variation in terms of forces it is
necessary to give up any timelike symmetry and squarely face the
time variability. The timelike Killing vector used earlier must be
replaced by the unit tangent vector to the world line of the
observer, $t$. In this rest frame the four-vector force representing
the accelerometer, $l^{\nu}$, will again have no zero component and
the tangent vector can be written as
\begin{equation}
t^{\mu}=f(x)\delta^{\mu}_{0},
\end{equation}
where$f(x)=(g_{00})^{-1/2}$. On account of this change of value of
$t^{\mu}$ the tidal-force expression becomes
\begin{equation}
F^{\mu}_{T}{=mf^2(x)R^\mu}_{0j0}l^{j},
\end{equation}
which differs from eq.(2) due to the extra $f^2(x)$ factor. Since
the Killing vector was not a unit vector it introduced a scaling
which appeared as a local Lorentz factor. This local Lorentz factor
was removed there by hand but here it is adjusted automatically.
Thus, while a re-scaled time appeared there, we are using proper
time here.

We need to take coordinates that are essentially synchronous
coordinates [12] but without the restriction $g_{00}=1$. Thus
$g_{0i}=0$. Now both ends of the accelertometer are spatially free,
i.e. both move and do not stay attached to some spatial point.
However, there is a "memory" of the initial time built into the
accelerometer in that the zero position is fixed then. Any change is
registered that way. Thus "time" behaves very differently from
"space". We must, therefore, use RNCs only for the spatial and not
for the temporal direction.

Thus using the block-diagonalised form of the metric with spatial
RNCs, eq.(8) can be written explicitly as
\begin{equation}
F^{\mu}_{T}=mf^2[\Gamma^{\mu}_{00,j}-\Gamma^{\mu}_{0j,0}
+\Gamma^{\mu}_{0j}\Gamma^{0}_{00}-
\Gamma^{\mu}_{0k}\Gamma^{k}_{0j}]l^{i}.
\end{equation}
This force, when extremised, can have no time component as
$\textbf{l}$ must be proportional to $\textbf{F}_{T}$ in that case.

Consider an observer in an FFF, equipped with an accelerometer by
which he can detect the tidal forces experienced by him. He can then
swivel the accelerometer till he gets a maximum reading on the dial.
Regarding eq.(9) as an eigenvalue equation with $\mu$ replaced by
$i$, the eigenvalue problem becomes
\begin{equation}
mf^2(\Gamma^{i}_{00,j}-
\Gamma^{i}_{0j,i}+\Gamma^{i}_{0j}\Gamma^{i}_{00}-\Gamma^{i}_
{0k}\Gamma^{k}_{0i})l^{j}=\lambda l^{i},
\end{equation}
where $\lambda$ is the eigenvalue and $l^{i}$ is the corresponding
eigenvector. Writing eq.(10) in matrix form we see that it has a
nontrivial solution only if $\lambda$ satisfies the cubic equation
\begin{equation}
\lambda^3+3a_1\lambda^2+3a_{2}\lambda+a_3=0,
\end{equation}
\begin{eqnarray}
\left\{\begin{array}{ll}
a_1=-(A+E+K)/3,\\
a_2=(AE+AK-BD-CG+EK-FH)/3,\\
a_3=-AEK+AFH+BDK-BFG+CEG-CDH;
\end{array}\right.
\end{eqnarray}
\begin{eqnarray}
\left\{\begin{array}{ll} A = mf^2[\Gamma^1_{00,1} -
\Gamma^1_{01,0}+\Gamma^1_{01}\Gamma^0_{00}-(\Gamma^1_{01})^2],\\
B=mf^2\Gamma^1_{00,2,},\quad C=mf^2\Gamma^1_{00,3},\quad D=mf^2\Gamma^2_{00,1,}\\
E=mf^2[\Gamma^2_{00,2}-\Gamma^2_{02,0}+\Gamma^2_{02}\Gamma^0_
{00}-(\Gamma^2_{02})^2], \\
F=mf^2 \Gamma^2_{00,3},\quad G=mf^2 \Gamma^3_{00,1},\quad H=mf^2
\Gamma^3_{00,2,}\\
K=mf^2[\Gamma^3_{00,3}-\Gamma^3_{03,0}
+\Gamma^3_{03}\Gamma^0_{00}-(\Gamma^3_{03})^2].
\end{array}\right.
\end{eqnarray}

Equation (11) can be solved to yield three roots and provide the
corresponding separation vector [13]. The general solution does not
provide much wisdom. As such we shall only make some observations
regarding the solution here. In the generic case there will be three
distinct eigenvalues. One of them, at least, will always be real. If
the other two are complex, the real value will give the required
$F^*_T$  and the corresponding eigenvector will give its direction.
If all three are real the maximum magnitude eigenvalue gives the
required tidal force, $F^*_T$, and the corresponding eigenvector
gives its direction. Further, if all three eigenvalues are real and
equal we have isotropy. If, due to extra symmetries, one of the
eigenvalues is zero we get a quadratic equation. In this case the
roots are always real. It is also possible that, due to further
symmetries, two roots are zero and we get a linear equation. Again
$F^*_T$ and the corresponding eigenvector are easily obtained.

\section{The Extended $\psi N$ Force}

Recall our definition of the relativistic analogue of the Newtonian
gravitational force. It is that quantity whose directional
derivative along the accelerometer, placed along the principal
direction, gives the extremised tidal force and which is zero in a
Minkowski space. Thus the $e\psi N$ force, $F^{\mu}$, satisfies the
equation
\begin{equation}
F^*_T=l^{\nu} F^{\mu}_{;\nu}.
\end{equation}
The fact that the zero component of the left side is zero does not
guarantee that the zero component of $F^{\mu}$ is zero. With the
appropriate gauge choice and using RNCs spatially, eq.(14) can be
written in a space and time break up as
\begin{equation}
l^{i}(F^{0}_{,i}+\Gamma^{0}_{ij}F^{j})=0,
\end{equation}
\begin{equation}
l^{j}(F^{i}_{,j}+\Gamma^{i}_{0j}F^{0})=F^{*i}_{T}.
\end{equation}
Notice that the expression in the brackets is essentially
$F^{*}_T\delta^{i}_{j}$ up to a scaling by the length of the
accelerometer.

A simultaneous solution of the above equations can be obtained by
taking the ansatz [13]
\begin{equation}
F^{0}=m[(\ln
A)_{,0}-\Gamma^{0}_{00}+\Gamma^{i}_{0j}\Gamma^{j}_{0i}/A]f^2,
\end{equation}
\begin{equation}
F^{i} = m\Gamma^{i}_{00}f^2,
\end{equation}
where
\begin{equation}
A=(\ln \sqrt{-g})_{,0},\quad g=det(g_{ij}),
\end{equation}
Now for these metrics
\begin{equation}
\Gamma^{0}_{00}=\frac{1}{2} g^{00}g_{00,0},\quad\Gamma^{i}_{00}=
-\frac{1}{2}g^{ij}g_{00,j,},\quad
\Gamma^{i}_{0j}=\frac{1}{2}g^{ik}g_{jk,0.}
\end{equation}
Thus the covariant form of the $e\psi N$ force, with the appropriate
choice of frame, is given by
\begin{equation}
F_{0}=m[(\ln Af)_{,0}+g^{ik}g_{jk,0}g^{jl}g_{il,0}/4A],
\end{equation}
\begin{equation}
F_{j}=m(lnf)_{,i}.
\end{equation}
The new feature of the $e\psi N$ force is its zero component. In
special relativistic terms, which are relevant for discussing forces
in a Minkowski space, the zero component of the four-vector force
corresponds to a proper rate of change of energy of the test
particle. Further, we know that in general an accelerated particle
either radiates or absorbs energy according as $dE/dt$ is greater or
less than zero. Thus $F_{0,}$ here, should also correspond to energy
absorption or emission by the background space-time. In fact we
could have separately anticipated that there should be energy
non-conservation as there is no timelike isometry. In that sense
$F_{0}$ gives a measure of the extent to which the space-time lacks
time isometry.

Another way of interpreting $F_{0}$ is that it gives a measure of
the change of the "gravitational potential energy" in the
space-time. In classical terms, neglecting this component of the
$e\psi N$ force would lead to erroneous conclusions regarding the
"energy content" of the gravitational field. Contrariwise, including
it enables us to revert to classical concepts while dealing with a
general relativistically valid treatment. It can be hoped that this
way of looking at energy in relativity might provide a pointer to
the solution of the problem of definition of mass and energy in GR.

\section{The $e\psi N$ Potentials}

It is clear that the gauge freedom could not be used to reduce the
gravitational potential to a single quantity for an arbitrary
space-time (as had been attempted earlier). Equations (17), (18) and
(19) provide five such quantities. The $e\psi N$ force, for
arbitrary space-times, can be expressed in the form of the
derivatives of two quantities. Now for our block-diagonalised
metrics
\begin{equation}
g^{ik}g_{ik,0}= -g^{ik}_{,0}g_{ik}.
\end{equation}
Thus the last term in eq.(21) can be reduced to
\begin{equation}
g^{ik}g_{jk,0}g^{jl}g_{il,0}=-g^{ij}_{,0}g_{ij,0}.
\end{equation}
Hence we can write
\begin{equation}
F_{0}=-U_{,0},\quad F_{i}=-V_{,j},
\end{equation}
with
\begin{equation}
U=m[\ln(Af/B)-\int(g^{ij}_{,0}g_{ij,0}/4A)dt],
\end{equation}
\begin{equation}
V=-m\ln f,
\end{equation}
where $B$ is a constant with units of time inverse, so as to make
$A/B$ dimensionless.

It is clear that $V$ is the generalization of the classical
gravitational potential and, for small variations from a Minkowski
space,
\begin{equation}
V\sim\frac{1}{2}m(g_{0}-1),
\end{equation}
which is the $\psi N$ potential. in fact the $e\psi N$ force for a
static spacetime is simply the $\psi N$ force with the Lorentz
factor adjusted. (Notice that it is reminiscent of the Kahler
potential.) It is in this sense that the $e\psi N$ potential is more
natural to use than the $e\psi N$ potential.

The quantity $U$ clearly represents a potential energy of the test
particle that contributes to its time variation. It is important
that the entire metric tensor (all ten components) is contained in
it. However, only the time-varying art of these components is
relevant. It is on account of this fact that the static metric has
only $g_{00}$ as the relevant potential. This is not the case for a
gravitational wave [13,14], for example. If $U$ is neglected, the
"Newtonian" institution will mislead us.

\section{Application to Some Geometries}

To better comprehend the significance of the $e\psi N$ force and
potential we consider two concrete examples. (In a separate paper we
discuss the application of this formalism to the problem of the
energy in gravitational waves [14]). The two examples are:\\
\par\noindent
a) the De Sitter metrics,\\
b) the Friedmann metrics.\\
\par\noindent
{\bf a) The De Sitter metrics}\\

The De Sitter metrics, with the observation point at the origin of
the polar coordinates, is defined by
\begin{equation}
ds^{2} = (I-r^{2}/D^{2})dt^{2}-(1-r^{2}/D^{2})^{-1}
dr^{2}-r^{2}-r^{2}d\Omega^{2},
\end{equation}
where $d\Omega^{2}=d\vartheta^{2}+sin^{2}\vartheta d\phi^{2}$ is the
solid angle element and $D=\sqrt{3/\Lambda}$ is the "radial distance
to the event horizon", where $\Lambda$ is the cosmological constant.
Then the tidal force is
\begin{equation}
F^{0}_{T}=F^{2}_{T}=F^{3}_{T}=0,\quad F^{1}_{T} = -ml^{1}/D^{2}
\end{equation}
and hence the maximum tidal force is simply $-ml/D^{2}$. This gives
the $e\psi N$ force
\begin{equation}
F^{0}=F^{2}=F^{3}=0,\quad F^{1}=-mr/D^{2}
\end{equation}
which is the usual force of "cosmical repulsion". The corresponding
$e\psi N$ potentials are then
\begin{equation}
U=0,\quad V=m\ln\sqrt{1-r^{2}/D^{2}}.
\end{equation}

Written in its exponentially expanding form the De Sitter metric
is
\begin{equation}
ds^{2} = d\tau^{2} - \exp[2\tau/D](dr^{2}+r^{2}d\Omega^{2}).
\end{equation}
In this form the tidal force is isotropic, but it has the same
maximal value as in the previous form. Here the $e\psi N$ force
becomes
\begin{equation}
F^{0}=F_{0}=-m/D,\quad F^{i}=0,
\end{equation}
and the corresponding $e\psi N$ potential is
\begin{equation}
U=-mr/D,\quad V=0
\end{equation}
The repulsive force has been replaced by a time-dependent "potential
energy", which again provides the red-shift. Here the universal
expansion has been built in, thus obviating the necessity for a
"force".

The $e\psi N$ formalism requires the use of the latter metric form,
eq.(33), instead of the former, eq.(29). The reason is hat the Lie
derivative of the separation vector, $\textbf{l}$, is not zero along
the unit time-like vector, $\textbf{t}$ given by
\begin{equation}
t^{\mu} = 1-r^{2}/D^{2})^{-\frac{1}{2}}\delta^{\mu}_{0}
\end{equation}
required for the former metric form. In physical terms, our
accelerometer is very small but non-negligible on the cosmological
scale. To the extent that it is non-negligible the time parameter is
re-scaled from one end of it to the other. In other words, in effect
the separation vector chosen does not lie in the purely spatial
direction, but has a temporal component. This is not so in the
latter case. Clearly, for cosmological purposes, we need to use a
synchronous coordinate system and must take $g_{00}=1$.\\
\par\noindent
{\bf b) The Friedmann metrics}\\

In the Friedmann cosmological models, due to the conservation of
mass energy, the energy density decreases with time as the
universe expands. We shall discuss only matter-dominated Friedmann
models
\begin{equation}
ds^{2}=dt^{2} - a^{2}(t)[d\chi^{2}+\alpha^{2}(\chi)d\Omega^{2}],
\end{equation}
where $\chi$ is the hyperspherical angle, $\sigma(\chi)$ is $\sin h
\chi,~\chi$ or $\sin \chi$ according as the model is open $(k=-1)$,
flat $(k=0)$ or closed $(k=1)$ and $a(t)$ is the corresponding scale
factor. The tidal force in this case is
\begin{equation}
F^{0}_{T} = 0,\quad F^{i}_{T}=-m\ddot{a}l^{i}/a,
\end{equation}
where dot "." denotes differentiation with respect to coordinate
time $t$, and not $s$. The maximal value is clearly $-m\ddot{a}l/a$.

For the flat Friedmann model, the extremal tidal force is
\begin{equation}
F^{*}_{T}=2ml/9t^{2}.
\end{equation}
Thus $F^{*}_{T0}\sim t^{-2}$ and hence $F^{*}_{T0}\rightarrow
\infty$ as $t \rightarrow 0$ and $F^{*}_{T0}\rightarrow 0$ as
$t\rightarrow \infty$

To discuss the general behaviour of the Friedmann models, let us,
first consider the extremal tidal force at the very early stages of
their evolution. For the early stages of the open (or closed)
Friedmann model, the scale factor will be
$a^{1/3}_{0}t^{2/3+\epsilon}$, where $\epsilon$ is a small positive
(or negative) quantity, and the extremal tidal force takes the form
\begin{equation}
F^{*}_{T}\sim(1-3\epsilon/2)F^{*}_{T0}.
\end{equation}
Thus the extermal tidal force will be less (or greater) than for the
flat case.

For the open Friedmann model at later times, the extremal value of
the tidal force is
\begin{equation}
F^{*}_{T^{-}}=4ml/a^{2}_{0}(cosh \eta-1)^{3}.
\end{equation}
Thus $F^{*}_{T} \sim t^{-2} $for large $t$, as in the case of
$F^{*}_{T0}$.

For the closed model, the extremal tidal force is
\begin{equation}
F^{*}_{T^{+}} = 4ml/a^{2}_{0}(1-\cos \eta)^{3}.
\end{equation}
Thus $F^{*}_{T^+}$ reaches a minimum value of
\begin{equation}
F^{*}_{T^+}=ml/2a^{2}_{0},
\end{equation}
at $\eta=\pi$ (i.e. at $t=a_{0}\pi/2$) and again becomes infinite at
$\eta=2\pi$ (or $t=a_{0}\pi$). The $e\psi N$ force, for the
Friedmann models, is simply
\begin{equation}
F^{0}=F_{0}=-m\ddot{a}/\dot{a},\quad F_{i}=0,
\end{equation}
and the corresponding $e\psi N$ potentials
\begin{equation}
U=m\ln(\dot{a}/b),\quad V=0,
\end{equation}
where $b$ is an arbitrary constant with unit of $\dot{a}$. For a
flat Friedmann model, eq.(4) yields
\begin{equation}
F^{0}F_{00}=m/3t,\quad F_(i)=0.
\end{equation}
The corresponding $e\psi N$ potentials are
\begin{equation}
U_{0} = \frac{1}{3}m\ln(t/T),\quad V=0
\end{equation}
for an appropriate choice of $b$. Thus $F_{0}$ is proportional to
$t^{-1}$ and hence $F_{0}$ goes to infinity as $t$ approaches zero
and it tends to zero when $t$ tends to infinity. Since $F_{0}$ is
positive, it corresponds to the energy absoprtion $[13]$ by the
background space-time. The comparison of the behaviour of the early
stages of the open and closed models with the flat, for the $e\psi
N$ force and potential, follows the tidal-force pattern exactly.

For the open Friedmann model at arbitrary times, the time component
of the $e\psi N$ force turns out to be
\begin{equation}
F_0-=2m/a_{0}sinh \eta(cosh \eta-1),
\end{equation}
and the corresponding $e\psi N$ potential is
\begin{equation}
U_{-}=m\ln[sinh \eta/(cosh \eta -1)].
\end{equation}
Hence$ F_{0-}$ goes as $t^{-1}$ for large $t$ as in the case of
$F_{00}$. Notice that $U_{-}\rightarrow 0$ as $t\rightarrow\infty$
whereas $U_{0}\rightarrow - \infty$ as $t\rightarrow\infty$. This
odd feature of the flat Friedmann model may indicate a problem with
our ansatz solution.

For the closed Friedmann universe there is a problem. The
Christoffel symbol appearing in eq.$(16)$ is zero, for this case,
when $F^{*}_{T}$ reaches a minimum value. According to the ansatz
used this gives an infinite $e\psi$N force at that instant. This is
clearly absurd. It was verified that obtaining the general solution
to eqs.$(15)$ and $(16)$ does not resolve this problem. However,
there is an arbitrariness in what we choose to call the "zero" of
the accelerometer. There is no \emph{a priori} reason to set it at
any particular value. We can then choose to set it at zero at the
phase of maxdimum expansion, $\eta=\pi$, so as to avoid the infinity
in the $e\psi N$ force. Using this resetting, the $e\psi N$ force
becomes
\begin{equation}
F^{0}=m\frac{(4+3\sin^{2}\eta+3\cos\eta+\cos^{3}\eta)}
{4a_{0}(1-\cos\eta)\sin\eta},\quad F^{i}=0
\end{equation}
This gives $F^{\mu}=0$ at the phase of maximum expansion of the
universe.

The $e\psi N$ potentials, here, are
\begin{equation}
U=m[\frac{1}{1-\cos \eta} + \frac{1}{4}\ln(1-\cos \eta)]/a_{0},\quad
V=0.
\end{equation} Hence, at the phase of maximum expansion the
gravitational potential energy is
\begin{equation}
U=m(1+\ln\sqrt{2})/2a_{0},\quad V=0.
\end{equation}

Generally, as $t\rightarrow0$, the $e\psi N$ force and potential for
the Friedmann metrics tend to infinity. In the closed model they
again become infinite at $\eta=2\pi$, i.e. $t = \pi a_{0}$.

\section{Conclusion}

It has been shown that the $e\psi N$-formalism, which had been
useful in providing some insights into the consequences of
relativity for a static space-time, \emph{can} be extended to
arbitrary space-times. The $e\psi N$ potential, in this case,
approximates the $\psi N$ potential for small variations from the
Minkowski space-time. The difference arises for nonstatic
space-times. The fundamentally new feature of the $e\psi N$ force is
its zero component. This does not contribute directly to the tidal
force, since the accelerometer has no zero co ponent in its own rest
frame, but it enters through the Christoffel symbols
$\Gamma^{0}_{ij}$ and $\Gamma^{i}_{0j}$. We find that this force can
be described in terms of two $e\psi N$ potentials in general.

We would like to reiterate, here, the importance of the choice of
frame. In doing so, to avoid possible confusion, we make it clear
that we refer not to the coordinate system but to the \emph{frame}.
All too often coordinates are regarded as merely labelling some
point in a manifold. While \emph{mathematically} correct, sight is
lost of the \emph{physical} point that in GR it includes the choice
of the reference frame. All physics is done with resect to some
frame and for any given experiment there will always be a preferred
frame. This argument does not militate against the principle of
general covariance but, rather, complements it. For any particular
process there \emph{must be} a frame in which it can be most simply
described. The point of general covariance is that physical laws,
generally, have no preferred frame which can be \emph{universally}
used for all purposes and in all circumstances. To discuss the
centrifugal force it is necessary to go into a corotating frame
[15]. Similarly, for the purpose of providing a $3+1$ split of the
space-time metric, we need to enter the freely falling rest frame,
which gives the $e\psi N$ force and potentials.

In view of the fact that an apparently internally consistent
expression for energy seems to be forthcoming from the $e\psi N$
formalism, it may be hoped that it will provide a handle to tackle
the problem of definition of mass and energy in GR. It could even be
hoped that such a definition of energy could ultimately lead to a
viable canonical quantization programme. However, there are still
some problems to be resolved. For one thing there is an
arbitrariness in setting the zero of the extremal tidal force, which
may be nontrivial. This zero of the extremal tidal force measured by
the accelerometer can be fixed according to the observer's choice.
We are using the RNCs spatially. This means that we have chosen the
three arbitrary constants, one in each spatial direction, to be
zero. Thus we have an arbitrariness in the zero setting of the
extremal tidal force which we can set by fixing the suitable
arbitrary constants.

There is another problem. The solution of eqs.(15) and (16) is not
unique but has been obtained by taking an ansatz. In fact this
solution is the "particular integral" part of the general solution.
There is another, "complementary", part to be obtained for the
general solution of the set of equations. It would be of interest to
solve the set of coupled, linear, inhomogenous, partial differential
equations given by eqs.(15) and (16). This solution would provide
further understanding of the definition of energy in GR given by the
$\psi N$ approach.

It is worthwhile to point out that the $e\psi N$ approach is
\emph{not} an alternative approach to GR. In fact it is very much a
part of GR. It is an attempt to understand the implications of this
theory better. The purpose is to recast the consequences in
Newtonian-"force" terms as an aid to our intuition. Other similar
attempts tend to give a "weak-field" or "linearized" effect. We, on
the other hand, use a method which allows us to consider
strong-field, non-linearized, effects. This would be \emph{vital} in
any attempt to "quantize relativity". (Here we would draw a
distinction between "quantizing gravity" and "relativity". The
latter is a theory of motion [2-4, 15].)

\vspace{0.5cm}

{\bf Acknowledgment}

\vspace{0.5cm}

We are very grateful to the Pakistan Science Foundation for
financial support during this research under the project
PSF/RES/C-QU/MATHS(16).

\vspace{0.5cm}

\textbf{REFERENCES}

\begin{description}

\item{[1]}  C.W. Misner, K.S. Thorne and J. A. Wheeler: \emph{Gravitation}
(W.H. Freemann, San Francisco, Cal., 1973).

\item{[2]} S.M. Mahajan, A. Qadir and P.M. Valanju: Nuovo Cimento \textbf{B65}
(1981)404.

\item{[3]} A. Qadir and J. Quamar: \emph{Proceedings of the Third Marcel
Grossman Meeting on general Relativity}, ed. Hu Ning (Science Press
and North Holland Publishing Co., 1983)p 189; Europhys. Lett.,
\textbf{2}(1986)432;\\
J. Quamar: Ph.D. thesis, Quaid-i-Azam University (1984).

\item{[4]} O. Gron: Phys. Lett. \textbf{A94}(1983)424; Gen. Relativ.
Gravit. \textbf{20}(1983)123;\\
A. Qadir: Phys. Lett. \textbf{A99}(1983)419.

\item{[5]} A. Qadir, J. Quamar and M. Rafique: Phys. Lett. \textbf{A109}(1985)92;\\
A. Qadir: Europhys. Lett. \textbf{2}(1986)427.

\item{[6]} A. Qadir and M. Rafique: \emph{Proceedings of the Fourth Marcel
Grossmann Meeting on General Relativity}, ed. R. Ruffini (Elsevier
Science Publishers, 1986)p 1597; Chin. Phys. Lett.
\textbf{3}(1986)189;\\
A. Qadir, M. Rafique and A.W. Siddiui: Chin. Phys. Lett.
\textbf{4}(1987)177; \\
A. Qadir: Chin. Phys. Lett. \textbf{4}(1987)289;\\
M. Rafique: Ph.D. thesis, Quaid-i-Azam University (1985);\\
Z.U. Mian, A. Qadir, J. Quamar and H.A. Rizvi: Commun. Theor. Phys.
\textbf{11}(1989)115.

\item{[7]} A. Qadir: \emph{Proceedings of the Third Regional Conference on
Mathematical Physics}, ed. F. Hussain and A. Qadir (World
Scientific, 1990)p481.

\item{[8]} S. Weinberg: Rev. Mod. Phys., \textbf{52}(1980)515;\\
A. Salam: Rev. Mod. Phys. \textbf{52}(1980)525;\\
S. L. Glashow: Rev. Mod. Phys. \textbf{52}(1980)539 and references
therein.

\item{[9]} S.W. Hawking and G.F.R. Ellis: \emph{The Large Scale Structure
of Spacetime} (Cambridge University Press, Cambridge, 1973).

\item{[10]} A.H. Bokhari and A. Qadir: \emph{Proceedings of the Fourth
Marcel Grossmann Meeting on General Relativity}, ed. R.Ruffini
(Elsevier Science Publishers, 1986)p 1632, 1635;\\
Z. Angew. Math. Phys. \textbf{3}(1986)189;\\
A.H. Bokhari: Ph.D. thesis, Quaid-i-Azam University (1986).

\item{[11]} B. Carter: \emph{Proceedings of the Marcel Grossmann Meeting on
General Relativity}, ed. by R. Ruffini (North Holland Publishing
Co., 1983)p 189.

\item{[12]} L.D. Landau and E.M. Lifshitz: \emph{The Classical Theory of
Fields} (Pergamon Press, 1975).

\item{[13]} M. Sharif: Ph.D. thesis, Quaid-i-Azam University (1991).

\item{[14]} A. Qadir and M. Sharif: Phys. Lett. \textbf{A167}(1992)331.

\item{[15]} M.A. Abramowicz, B. Carter and J.P. Lasota: Gen. Relativ.
Gravit. \textbf{20}(1998)1173;\\
M.A. Abramowicz: Mon. Not. R. Astron. Soc. \textbf{245}(1990)733.

\end{description}
\end{document}